\documentstyle[12pt]{article}
\newcommand{\be}{\begin{equation}}
\newcommand{\ee}{\end{equation}}
\newcommand{\rn}{${\cal R}^n$~}
\newcommand{\Rn}{$RanD(M,d\mu)$~}
\begin{document}
\begin{center}
{\bf \large Space--time discretization: the way to the fundamental element
without a determined form.}
\vskip10mm
{ S.B. Afanas'ev}

{\bf \it St.--Petersburg, Russia}
\end{center}
\begin{abstract}
The concept of the random discretization of the space--time is suggested.
It is the way to consistent compatible synthesis of quantum and
relativistic principles and principle of geometrization. The basic idea of
this concept is physical reality of the finite sizes fundamental element of
the quantized space--time. The flat space--time with random
discretization is described
as the probability measure space with the set of all possible discretizations
of the flat continual space--time as the set of points. The probability
measure can depend on the geometric parameters of discretizations (a number
of regions of a discretization, their volumes, areas etc.). In this concept
the fundamental length can be defined as the average value of the linear size
of a fundamental element. In this concept the "particle" quantum and the
space-time quantum are identical.
\end{abstract}

The synthesis of relativistic, quantum and geometrization principles is one
of the central problem of modern physics. Below this set of principles
(quantum,
relativistic and principle of geometrization) is conventionally
named QRG
principles. The approaches to their consistent compatible realizing are found
both on the way of the solution of the quantum gravity problem and
quantization of the space--time~\cite{quant} and in the creation of the
united theory of
fundamental interactions and the geometrized particle theory.
In the last theories
(in particular, in the superstring theory~\cite{gsw})
the particles are represented as
the excited states of some extended objects. In this work an approach, in which
the particles can be described as
the excited states of $n$--dimensional fundamental elements
of the space--time, is suggested. This approach is based on the representation
of discrete structure of the quantized space--time (QST) as the set of finite
size fundamental elements. We suppose that
these fundamental elements have the property
of physical reality. In other words, the physical space--time is
the one with the discrete structure, but not continual
space--time that consist of points, i.e. fundamental elements without sizes
and structure. The continual space--time is the limiting case of the
discrete space--time with the
zeroth value of the fundamental length and has the
sense of mathematical abstract.

The consideration of the space--time fundamental element as the physical
reality element and the principles of quantum description of microobjects
necessarily lead to abandonment of determined form and sizes of a fundamental
element (since they are variables describing quantum object and cannot
have exactly determined values)
and representation of QST as the probability measure space with the
set of all possible discretizations of the continual space--time as the set of
points. Therefore this concept is properly named the concept of the random
discretization of the space--time. At first glance this concept is more
compatible consistent realizing of QRG principles than both the known
approaches to the space--time discrete structure description and the
existed geometrized theories of particles. Thus in the superstring theory
the "particle" quantum (i.e. all particles are excited states of this quantum)
is not identical to the quantum of the space--time, and the string
as quantum geometrical object can be defined as the space quantum only.
In the suggested concept the "particle" quantum
is the fundamental element of the quantized space--time, or the quantum
of the space--time. On the other hand, this concept contain many other
discrete space--time models~\cite{disc}
(the lattice space--time, the Regge simplicial
space--time etc.) as special cases of investigated object with the
special choices of the probability measure.

In this work the basic ideas and representations of the concept of the
random discretization are discussed. They are concerned with the semiclassical
method of description of discrete structure of the space--time. The questions
about nature of particles as the excitations of QST is briefly discussed in
the conclusion. In this work the case of the space--time with the average
values equal over all discretized manifold is analyzed.

Consider the semiclassical description of the discrete structure of QST
in the framework of the concept of the random discretization. Below the
geometric base of semiclassical description of the flat space--time (or
the space) is discussed. Obviously, difference between the space--time and
the space is not important as long as the metric relations are not
considered.

The manifold of the flat space
with random discretization (briefly - random discretized flat manifold)
is the probability measure space
with the set of all possible discretizations of $M\subset$\rn as the
set of points. The probability measure
defined on this set can depend on the scalars of discretizations, i.e.
a number of regions of a discretization, their
volumes, areas of surface etc. Below we suppose that the measure on the
set of discretizations can be defined.
Correct definition of the probability measure on the set
of all \rn--~discretizations has some difficulties, and
the flat space with random discretization can be defined as the set of all
random discretized flat manifolds (or representative set of manifolds, for
example, all spaces with the set of discretizations of open sphere).
Denote the set
of discretizations $D(M)$ and the discrete space with random discretization
\Rn, where $M$ is the discretized manifold of \rn, $d\mu$ is
the probability measure.
The probability measure $d\mu$ is represented in the form
\be
d\mu = \frac{1}{I} \, \cdot \mu_{p} d\sigma,
\ee
where $d\sigma$ is the measure on the set of discretizations,
$\mu_{p}$ --- the factor characterized non-equality of the probabilities
of discretizations,
$I = \int \limits_{D(M)} \mu_{p} \,\,d\sigma$
The sense of notation $\mu_{p}$ is following: different values
of probability density of discretizations are caused by the physical
properties of excitations of the space--time, and it can be says that
$\mu_{p}$ is the physical factor of the probability measure.

Below only the measures dependent on the number of regions $N$, their
volumes $\{V_i\}^{N}_{i=1}$ and their surface areas $\{S_i\}^{N}_{i=1}$ are
considered. In the general case $\mu_{p}$ is
represented as the sum of series
\be
\label{mes}
\mu_{p} = \sum A_{\alpha \{\beta\} \{\gamma\}} N_{\alpha}
V_1^{\beta_1} ... V_N^{\beta_{N}} S_1^{\gamma_1}... S_N^{\gamma_{N}}
\ee

The average values of $N$, $V$ and $S$ are defined in the following manner:
\be
<N> = \int \limits_{D(M)} N d \mu
\ee
\be
\label{V}
<V> = \int \limits_{D(M)} {\bar V}^{\{l\}} d \mu^{\{l\}}
\ee
where ${\bar V}^{\{l\}} = \frac{1}{N} \sum_{i=1}^{N} V_i^{\{l\}}$,
$\{l\}$ is identified the discretizations
\be
\label{S}
<S> = \int \limits_{D(M)} {\bar S}^{\{l\}} d \mu^{\{l\}}
\ee
where ${\bar S}^{\{l\}} = \frac{1}{N} \sum_{i=1}^{N} S_i^{\{l\}}$
Thus the average values of $V$ and $S$ are calculated by double--averaging:
over the regions of a discretization and over discretizations.

Consider also the subset of the probability measures from the set (\ref{mes})
that are represented in the form:
\be
\mu_{p} = f ( \sum_{i=1}^{N} A_{\alpha \beta \gamma} N^{\alpha} V_i^{\beta}
S_i^{\gamma})
\ee
Obviously, basic
interest is caused by the measures for which the average values $<N>$,
$<N>$ and $<S>$ are finite.

It can be supposed that these measures must satisfy following conditions
\be
\label{c1}
\lim_{N \to 1} \mu_{p} = \mu_1
\ee
\be
\label{c2}
\lim_{N \to \infty} \mu_{p} = 0
\ee

These conditions are necessary because both the finiteness of values of
$\mu_{p}$
for discretizations with the small number of regions $N$ and the zeroth limit
of $\mu_p$ with $N \to \infty$ are required for
the finiteness of integrals by type
\be
<N> = \int N d \mu
\ee

The problem, are the conditions (\ref{c1}, \ref{c2})
sufficient for the
finiteness of values $<N>$, $<V>$ and $<S>$, is open. It is not excluding that
the finiteness of values $<S>$ requires the satisfaction of special conditions
for values of the probability measure for discretizations that contain
the regions with the large values S. It can be suggested several dependencies
of $\mu_{p} (V,S,N)$ that are satisfied the conditions (\ref{c1}) and
(\ref{c2})
\be
\mu_{p} = C\, exp(-\,A \, \sum_i S_i)
\ee
\be
\mu_{p} = C\, exp(-\,A \, \sum_i \frac{S_i}{V_i})
\ee
\be
\mu_{p} = C\, exp(-\,A \, \sum_i \frac{S_i}{V_i} -B \sum_i \frac{V_i}{S_i})
\ee

Besides the problem of dependence of average values $<V>$, $<S>$ and
$<N>$ on the size of a discretized manifold has the deep sense.
Obviously, this dependence must
satisfy some requirements for the measures interesting for the description
of the physical space--time. It seems likely that this problem is connected
with the conformal invariance in the investigated discrete space.
At first glance the condition of independence (or weak dependence) of average
values of $V$ and $S$ on the continual manifold size is most reasonable
condition.

Thus the considered probability measure space is realizing of discrete
space (space--time) structure with the finite value of the fundamental
length (with the supposition of existence of the set of probability
measures that give the finite values of $N$, $V$ and $S$). It is noted that
earlier investigated concepts of space--time discretization are the special
cases of considered object. Thus the lattice space--time is the special
case of \Rn with the following choice of the probability measure:
$\mu = \mu_0$ for discretizations with the lattice as the set of boundaries
of regions, and $\mu = 0$ for other discretizations. The Regge simplicial
space--time is the special case of considered space with the probability
measure different from zero for discretizations by simplexies.

In this concept the fundamental length has the meaning some average linear
size of regions of discretizations. It can be suggested three geometric
parameters by this type:
\be
l_I = <2 n\, \frac{V}{S}>
\ee
\be
l_{II} = <V^{1/n}>
\ee
\be
l_{III} = <V>^{1/n}
\ee
where $n$ is the dimensionality of the space--time, and the average values
are defined by the method of double--averaging (see (\ref{V}), (\ref{S})).

Average values of the geometrical variables can
be calculated with using of the construction that is analogous to the
continual (functional) integral. This construction can be considered as
the generalization
of the Feynman integral over trajectories and the string world surfaces
integral.

In conclusion, it is some words about development of considered concept.
Considered method of discretization of the flat space--time is the first step
to realizing of synthesis of relativistic, quantum and geometrization
principles. This method of discretization of the space--time gives the way
to the introducing of mathematical operations on the set of regions of
discretizations (different from the continual space--time operations).
This way can be more consistent realizing of three basic principles than
the string and brane theories and other concepts of the space--time
discretizations by the following causes:

1) This space--time is compatible
quantized (in this context, disretized) automatically
(connection with the algebraic problem of space-time
quantization see below), and the continual space--time
don't play the role of method of space--time description as it is in the
superstring theory;

2) This space--time description method allow to introduce the operations
on the set of fundamental elements (regions of discretizations), and
principally property of
relativistic invariance can be formulated in terms of quantized (discrete)
space--time;

3) Particles can be considered as excited states of fundamental elements
of the discrete space--time, and different particles are described
by the different dependence of the probability measure on
the parameters of discretizations. Last conclusion is in the agreement with
the form of general construction of the functional integral, in which
$\mu_{p}$ is coincided with factor $exp(-S)$, where $S$ is an action;

4) In this concept the "particle" quantum is identical to the space-time
quantum. Thus this concept meet the principle of minimal number of basic
object on the most fundamental level.

It is noted the following problems for the development of this concept:

1) Formulation of \Rn--~coordinate--~dependent probability space (space--
time) in the framework of semiclassical description considered in this
article;

2) Introduction of the mathematical operations on \Rn. This approach allow
to describe space--time in internal terms of invariance and transformations;

3) Research of the random discretization of the curved
space--time. In this concept all tensors are the random ones, and equations
and equalities of curved discrete space--time have the probability sense;

4) Connection between the discrete structure of the space--time and the
algebraic description of the quantized space--time (commutation relations,
introducing of operators of creation and destruction etc.).
The solution of this problem
is required introduction of the fundamental elements of quantized space--time
as the elementary units of not \rn, but the all investigated probability
measure space, i.e. introduction of elementary units of the set of all
discretizations. In this approach operators of QST are
described the fundamental elements of the one;

5) Relation between calculations over the set of discretizations and the
ones over the topology (set of all possible subsets of \rn).
This problem is connected with the problem of
correspondence of the probability measures defined on the set of
discretizations and on the topology~\cite{bal};

6) Formulation of the compatible quantum description of the discrete
structure of the space-time, i.e. definition of the wave function on the
set of discretizations and finding of basic equations and fundamental
properties of this wave function.

\end{document}